\documentclass[aps,preprint, floatfix]{revtex4}

\usepackage{graphicx,color}
\usepackage{psfrag}
\usepackage{amssymb}
\usepackage{eso-pic}
\usepackage[dvips,letterpaper]{geometry} 
\usepackage{amsmath}

\begin{document}

%
%
%
%
\def\oti{{\otimes}}
\def\lb{ \left[ }
\def\rb{ \right]  }
\def\tilde{\widetilde}
\def\bar{\overline}
\def\hat{\widehat}
\def\*{\star}
\def\[{\left[}
\def\]{\right]}
\def\({\left(}		\def\BL{\Bigr(}
\def\){\right)}		\def\BR{\Bigr)}
	\def\BBL{\lb}
	\def\BBR{\rb}
%
%
\def\zb{{\bar{z} }}
\def\zbar{{\bar{z} }}
\def\frac#1#2{{#1 \over #2}}
\def\inv#1{{1 \over #1}}
\def\half{{1 \over 2}}
\def\d{\partial}
\def\der#1{{\partial \over \partial #1}}
\def\dd#1#2{{\partial #1 \over \partial #2}}
\def\vev#1{\langle #1 \rangle}
\def\ket#1{ | #1 \rangle}
\def\rvac{\hbox{$\vert 0\rangle$}}
\def\lvac{\hbox{$\langle 0 \vert $}}
\def\2pi{\hbox{$2\pi i$}}
\def\e#1{{\rm e}^{^{\textstyle #1}}}
\def\grad#1{\,\nabla\!_{{#1}}\,}
\def\dsl{\raise.15ex\hbox{/}\kern-.57em\partial}
\def\Dsl{\,\raise.15ex\hbox{/}\mkern-.13.5mu D}
%
%
\def\ga{\gamma}		\def\Ga{\Gamma}
\def\be{\beta}
\def\al{\alpha}
\def\ep{\epsilon}
\def\vep{\varepsilon}
\def\la{\lambda}	\def\La{\Lambda}
\def\de{\delta}		\def\De{\Delta}
\def\om{\omega}		\def\Om{\Omega}
\def\sig{\sigma}	\def\Sig{\Sigma}
\def\vphi{\varphi}

%
%
\def\CA{{\cal A}}	\def\CB{{\cal B}}	\def\CC{{\cal C}}
\def\CD{{\cal D}}	\def\CE{{\cal E}}	\def\CF{{\cal F}}
\def\CG{{\cal G}}	\def\CH{{\cal H}}	\def\CI{{\cal J}}
\def\CJ{{\cal J}}          \def\CK{{\cal K}}	\def\CL{{\cal L}}
\def\CM{{\cal M}}	\def\CN{{\cal N}}	\def\CO{{\cal O}}
\def\CP{{\cal P}}	\def\CQ{{\cal Q}}	\def\CR{{\cal R}}
\def\CS{{\cal S}}	\def\CT{{\cal T}}	\def\CU{{\cal U}}
\def\CV{{\cal V}}	\def\CW{{\cal W}}	\def\CX{{\cal X}}
\def\CY{{\cal Y}}	\def\CZ{{\cal Z}}

\def\rvac{\hbox{$\vert 0\rangle$}}
\def\lvac{\hbox{$\langle 0 \vert $}}
\def\comm#1#2{ \BBL\ #1\ ,\ #2 \BBR }
\def\2pi{\hbox{$2\pi i$}}
\def\e#1{{\rm e}^{^{\textstyle #1}}}
\def\grad#1{\,\nabla\!_{{#1}}\,}
\def\dsl{\raise.15ex\hbox{/}\kern-.57em\partial}
\def\Dsl{\,\raise.15ex\hbox{/}\mkern-.13.5mu D}
%
%
%
\font\numbers=cmss12
\font\upright=cmu10 scaled\magstep1
\def\stroke{\vrule height8pt width0.4pt depth-0.1pt}
\def\topfleck{\vrule height8pt width0.5pt depth-5.9pt}
\def\botfleck{\vrule height2pt width0.5pt depth0.1pt}
\def\Zmath{\vcenter{\hbox{\numbers\rlap{\rlap{Z}\kern
0.8pt\topfleck}\kern 2.2pt
                   \rlap Z\kern 6pt\botfleck\kern 1pt}}}
\def\Qmath{\vcenter{\hbox{\upright\rlap{\rlap{Q}\kern
                   3.8pt\stroke}\phantom{Q}}}}
\def\Nmath{\vcenter{\hbox{\upright\rlap{I}\kern 1.7pt N}}}
\def\Cmath{\vcenter{\hbox{\upright\rlap{\rlap{C}\kern
                   3.8pt\stroke}\phantom{C}}}}
\def\Rmath{\vcenter{\hbox{\upright\rlap{I}\kern 1.7pt R}}}
\def\Z{\ifmmode\Zmath\else$\Zmath$\fi}
\def\Q{\ifmmode\Qmath\else$\Qmath$\fi}
\def\N{\ifmmode\Nmath\else$\Nmath$\fi}
\def\C{\ifmmode\Cmath\else$\Cmath$\fi}
\def\R{\ifmmode\Rmath\else$\Rmath$\fi}

\def\barray{\begin{eqnarray}}
\def\earray{\end{eqnarray}}
\def\beq{\begin{equation}}
\def\eeq{\end{equation}}

\def\n{\noindent}

\def\Tr{\rm Tr} 
\def\xvec{{\bf x}}
\def\kvec{{\bf k}}
\def\kvecp{{\bf k'}}
\def\omk{\om_{\kvec}} 
\def\2pi2{(2\pi)^2}
\def\ket#1{|#1 \rangle}
\def\bra#1{\langle #1 |}
\def\adag{a^\dagger}
\def\rme{{\rm e}}
\def\Im{{\rm Im}}
\def\pvec{{\bf p}}
\def\fermiS{\CS_F}
\def\cdag{c^\dagger}
\def\adag{a^\dagger}
\def\bdag{b^\dagger}
\def\vvec{{\bf v}}
\def\vac{|0\rangle}
\def\gradvec{\vec{\nabla}}
\def\psidag{\psi^\dagger} 
\def\up{\uparrow}
\def\down{\downarrow}
\def\smallhalf{{\textstyle \inv{2}}}
\def\smallsqrt{{\textstyle \inv{\sqrt{2}}}}
\def\rvec{{\bf r}}
\def\avec{{\bf a}}
\def\pivec{{\vec{\pi}}}
\def\dim#1{\lbrack\!\lbrack #1 \rbrack\! \rbrack }
\def\angstrom{{{\scriptstyle \circ} \atop A}     }
\def\AA{\leavevmode\setbox0=\hbox{h}\dimen0=\ht0 \advance\dimen0 by-1ex\rlap{
\raise.67\dimen0\hbox{\char'27}}A}

\def\Im{{\rm Im}}
\def\Re{{\rm Re}}

\def\Li{{\rm Li}}

\def\dim#1{{\rm dim}[#1]}

\def\ep{\epsilon}

\def\free{\CF}

\def\Fhat{\digamma}

\def\CI{\mathcal{I}}

\def\eF{\epsilon_F}

\def\pvec{{\bf p}}

\def\Kvec{{\bf K}}

\def\kappavec{\vec{\kappa}}

\def\ihat{\hat{i}}
\def\jhat{\hat{j}}

\def\cutoff{\delta \kappa}

\title{Superconductivity in the two-dimensional Hubbard model based on the
exact  pair potential}
\author{ Andr\'e  LeClair}
\affiliation{Newman Laboratory, Cornell University, Ithaca, NY} 
\affiliation{Centro Brasileiro de Pesquisas F\'isicas,  Rio de Janeiro}.

\begin{abstract}

We analyze solutions to a  superconducting gap equation based on the   two-dimensional
Hubbard model with nearest and next-to-nearest neighbor hopping.   The Cooper
pair potential can be calculated exactly and expressed in terms of  standard elliptic functions.
The Fermi surfaces at finite temperature and chemical potential are also calculated 
based on the exact two-body S-matrix of the Hubbard model using the formalism we
recently developed\cite{SHubbard},  which allows variation of  hole doping.  
 The resulting solutions to the gap equation 
are strongly anisotropic,  namely largest in the anti-nodal direction, and zero in the nodal directions
of the Brillouin zone,  but not precisely d-wave. 
 For $U/t = 13$ and  $t' /t =-0.3$, appropriate to BSCO,   and a
physically natural choice for the cut-off,  our self-contained  analytic calculations  yield  
$\Delta_{\rm anti-nodal} /t   \approx 0.06$ and  maximum  $T_c /t \approx 0.04$  at  optimal hole doping
$h=0.15$.     For phenomenological fits to the Fermi surfaces for cuprates,  we obtain 
the comparable value $T_c/t = 0.03$ at optimal doping,  both in good agreement with experiments. 
The superconducting gap is non-zero for all hole-doping $h<0.35$ and increases all the way down
to zero doping,  suggesting that it evolves  smoothly into the pseudogap.

\end{abstract}

\maketitle

\section{Introduction}

The microscopic physics underlying  high $T_c$ superconductivity in the cuprates is
believed to be purely electronic in origin,   and strongly correlated electron  models such
as the two-dimensional  Hubbard model have been proposed to describe it\cite{Anderson}.   
A partial list of  more recent articles addressing the existence of superconductivity in the Hubbard model 
is \cite{Maier,MonteCarlo,Raghu},  and references therein. 
The Hubbard model simply describes electrons hopping on a square lattice subject to 
strong, local, coulombic  {\it repulsion}.    Since it is known that the condensed charge carriers
have charge $2e$,  and thus some kind of Cooper pairing is involved,  a central question has become ``What provides
the glue that pairs the electrons?''.    This  is especially puzzling since the underlying
bare interactions are repulsive.    The  situation is completely different in the 
Bardeen-Cooper-Schrieffer (BCS) theory of ordinary superconductors,  where the
attractive glue is provided by the interaction of the electrons with the lattice phonons\cite{BCS}.

Since the Mott-insulating anti-ferromagnetic phase at half-filling is well understood,
a large portion of the theoretical literature starts here  and attempts to understand  how doping
``melts''  the anti-ferromagnetic order,  and how the resulting state can become 
superconducting.    This has proven to be quite challenging,  perhaps in part due to
the fact that anti-ferromagnetic order is spatial,  whereas superconducting order is in
momentum space.  Consequently this has 
led to many interesting and in cases exotic ideas,   however  the central question, 
 ``What is the glue?'',   and how it arises from strongly coupled physics,  
remains unclear.    
 (For a review and other refereces,  see \cite{Wen}.)   
  This suggests that it may be more fruitful to 
begin on the overdoped side,  far away from any competing anti-ferromagnetic order,  in order
to understand the attractive mechanism in a pure form.    
Here the density is perhaps low enough that one can treat the model as a gas,  with 
superconductivity arising as a condensation of Cooper pairs  as in the BCS theory,
and we will adopt this point of view in the present work.     The observation made 
in \cite{SHubbard}  now comes to bear on the problem:   multi-loop quantum corrections
to the scattering of Cooper pairs can actually lead to effectively attractive interactions,
even though the bare model was defined with repulsive interactions.    In the work \cite{SHubbard}, 
the focus was on the thermodynamics at finite temperature and chemical potential,  
and evidence was presented 
 for instabilities toward the formation of new phases as the temperature was 
lowered.    However our purely thermodynamic 
 formalism was unable to probe the nature of the ground states of
these potentially new phases.   The present work attempts to complete the picture.   
Namely,   we explore how the attractive interactions described in \cite{SHubbard} 
can lead to superconductivity,  and what its basic properties are.

Our starting point will be the BCS theory,  but specialized to the Hubbard model.    
The original Cooper argument\cite{Cooper} is quite robust,  and shows that 
any attractive interactions near the Fermi surface lead to a pairing instability.      
We thus assume that the BCS construction  of the ground state 
goes through,   leading to the well-known gap equation\cite{BCS}:
\beq
\label{gapeqn}
\Delta (\kvec )  = -   \int  \frac{d^2 \kvec'}{\2pi2}    ~ V(\kvec, \kvec')    \frac{\Delta (\kvec' )}{2 E(\kvec ')}
\, \tanh (E(\kvec')/2T), 
~~~~E(\kvec ) \equiv     \sqrt{\xi  (\kvec)^2  +  \Delta (\kvec )^2 }
\eeq
Here,  $\Delta$ is the energy gap,   $E(\kvec )$ is the energy of excitations above the ground state, 
and $T$ is the temperature.   
We will later make some favorable checks on the approximations that lead to the above equation.   
In the above gap equation,   $V$ represents the residual interaction of Cooper pairs,  
and we will refer to it as the (Cooper) pair potential.    The main new input is that we compute
$V$ from the Hubbard model,  including quantum corrections,  and show that it has attractive regions
in the Brillouin zone.         
The other ingredient is $\xi (\kvec)$,   which represents normal state quasi-particle energies near the Fermi surface, 
where $\xi =0$ at the Fermi surface.    This can be identified with the ``pseudo-energy'' 
in the thermodynamic approach described in \cite{SHubbard}.     
These two ingredients lead to a self-contained analysis of the solutions of the above gap equation
based entirely on analytic calculations carried out in  the Hubbard model.

An outline of the  sequel,  along with a summary of our results,  goes as follows.  
In the next section we describe our conventions for the Hubbard model, with hopping 
strengths $t,t'$ and the repulsive coupling  $U>0$, all with units of energy.  
The pair potential $V$ is calculated in section III by summing multi-loop Feynman diagrams;
the final result is expressed in terms of elliptic functions.   It is demonstrated that,  rather surprisingly, 
when $U/t$ is large enough,  there opens up a region of attractive interactions,  i.e. negative $V$, 
near the half-filled Fermi surface.     In section IV,  the method developed in \cite{PyeTon,SHubbard} 
for the thermodynamics is reviewed,  and Fermi surfaces are calculated as a function of doping
for $U/t = 13$ and $  t'/t = -0.3$,  appropriate to the cuprate BSCO.       The results in these sections III, IV constitute the main inputs 
for the study of the solutions of the gap equation,  which is carried out in section V.   
The values we compute for the gap $\Delta$ and $T_c$ are in reasonably good agreement with
experiments.   The gap is anisotropic,  in that it vanishes in the nodal directions and is largest
in the anti-nodal,  however it is not precisely of d-wave form.  
Our  solutions to the gap equation persist to arbitrarily low doping,  and we propose that
they evolve into the so-called pseudogap,   in accordance with recent experiments.     In section VI,  
we repeat the analysis of the gap equation using a phenomenological fit to the Fermi surfaces.

\section{The Hubbard model gas}

The Hubbard model describes fermionic particles  with spin,   hopping between the sites
of a square lattice, subject to strong local coulombic repulsion.  
The lattice hamiltonian is 
\beq
\label{Hublat}
H= - t \sum_{<i,j>, \alpha = \up ,\down} \(  c^\dagger_{\rvec_i , \alpha} 
c_{\rvec_j , \alpha}   \)
 - t' \sum_{<i,j>', \alpha = \up ,\down} \(  c^\dagger_{\rvec_i , \alpha} 
c_{\rvec_j , \alpha}   \) 
 +  U \sum_{\rvec} n_{\rvec\up} n_{\rvec \down} 
\eeq
where $\rvec_{i,j} ,  \rvec$ are sites of the lattice,  $<i,j>$ denotes
nearest neighbors,  $n = c^\dagger c $ are densities,   and $c^\dagger, c$ satisfy canonical anti-commutation
relations.     
We have also included a next to nearest neighbor hopping term $t'$, 
since it is not difficult to incorporate  into the formalism,  and it is known to be  non-zero
for high $T_c $ materials.    

The free part of the  hamiltonian, i.e. the hopping term,   is easily diagonalized:
\beq
\label{freeH}
H_{\rm free} =  \int d^2 \kvec ~  \omega_\kvec \sum_\alpha c^\dagger_{\kvec, 
\alpha} c_{\kvec, \alpha} 
\eeq
with  the free 1-particle energy
\beq
\label{omeg}
\omega_\kvec =  - 2 t \(  \cos( k_x a)  + \cos (k_y a ) \)   -4 t' \cos( k_x  a)  \cos (  k_y  a)
\eeq
where  $t$ taken to be   positive. 
In the sequel it is implicit
that $\kvec$ is restricted to the first Brillouin zone, 
$-\pi/a \leq k_{x,y} \leq \pi/a$,   where $a$ is the lattice spacing.

Since the quartic interaction is local,   
we introduce the two continuum  fields $\psi_{\up, \down}$ 
and the action
\beq
\label{action}
S = \int d^2 \rvec \,  dt \( \sum_{\alpha= \up, \down}  i  \,  \psi^\dagger_\alpha \d_t \psi_\alpha -
\CH \)
\eeq
where $\CH = \CH_{\rm free}  + \CH_{\rm int}$ is the hamiltonian density.  
The field has the following expansion characteristic of a non-relativistic theory since it only involves
annihilation operators,
\beq
\label{field}
\psi_\alpha (\rvec ) = \int \frac{d^2 \kvec}{(2\pi)^2} ~ 
c_{\kvec, \alpha}  \, e^{i \kvec \cdot \rvec} 
\eeq
and satisfies at equal times 
$
\{ \psi_\alpha (\rvec ) , \psi^\dagger_{\alpha'} (\rvec') \} = \delta(\rvec - \rvec') \delta_{\alpha, \alpha'}
$.   
Since we have represented sums over lattice sites $\rvec$ as $\int  d^2 \rvec /a^2$,  
where $a$ is the lattice spacing,  
$c_\rvec  =  a  \psi (\rvec )$.   
The interaction part of the hamiltonian is approximated as a continuum integral with
density 
\beq
\label{hint} 
\CH_{\rm int} =  \frac{u}{2}   ~  \psi^\dagger_\up \psi_\up 
\psi^\dagger_\down \psi_\down 
\eeq
where $u = 2  U a^2$. 
Formally,  the free part of the hamiltonian density is 
$
\CH_{\rm free}   =  \sum_{\alpha = \up , \down}  \psi^\dagger_\alpha   W(\gradvec_{\rvec} ) 
\psi_\alpha  
$, 
where $W$ is the differential operator $W =  \omega (\kvec \to -i \gradvec_{\rvec} )$, 
and is thus non-local.       However this non-locality does not obstruct the solution of the 
model since the free term can be diagonalized exactly.   
 The model can now be treated as a quantum
fermionic gas,  where the only effect of the lattice is in the free particle energies
$\omega_\kvec$.    
 
  The field $\psi$ has dimensions of  inverse length,
and the coupling $u$ units of ${\rm energy}  \cdot {\rm length}^2 $.   
In the sequel we will scale out the dependence on $t$ and the lattice spacing $a$,  
and physical quantities will then depend on the dimensionless coupling
\beq
\label{gdef} 
g =   \frac{u}{a^2 t}   = \frac{2U}{t}
\eeq
Positive $g$ corresponds to
repulsive interactions.    Henceforth,  all  energy scales,   in particular,  the single particle energies
$\omega_\kvec$,
the gap $\Delta$,  
temperature,  and chemical potential,  will be implicitly in units of $t$.    

For both  cuprates  LSCO and BSCO,   $t \approx 0.3 ev  \approx 3000K$,  $U/t \approx 13$,  and 
 $t'/t$   approximately equals $-0.1$ and
$-0.3$ respectively.    Therefore,  for most of the detailed analysis  below, 
 we  set $g=26$ and  $t'/t = -0.3$ appropriate to BSCO.


\section{The Cooper pair potential  $V$.}

The kernel $V(\kvec, \kvec' )$  in the gap equation (\ref{gapeqn}) represents the
residual interaction of Cooper pairs of momenta $(\kvec, - \kvec)$ and 
$(\kvec' , - \kvec' )$.   It is related to the following matrix element of the interaction 
hamiltonian:
\beq
\label{VHint}
V(\kvec , \kvec' )  =  \int d^2 \rvec  ~ 
\langle \kvec' \up, - \kvec' \down |   \CH_{\rm int} (\rvec )  |  \kvec \up,  - \kvec \down \rangle
\eeq
To lowest order,   $V$ is momentum independent:  $V = g/2$.    

\begin{figure}[htb] 
\begin{center}
\hspace{-15mm} 
\psfrag{k1}{$\kvec_1$}
\psfrag{k2}{$\kvec_2$}
\psfrag{k3}{$\kvec_3$}
\psfrag{k4}{$\kvec_4$}
\psfrag{p}{$p$}
\psfrag{p12p}{$p_{12} - p$}
\psfrag{p34p}{$p_{34} - p$}
\psfrag{1}{$1$}
\psfrag{2}{$2...$}
\psfrag{m}{$m$}
\psfrag{a}{$m+1..$}
\psfrag{N}{$N$}
\includegraphics[width=13cm]{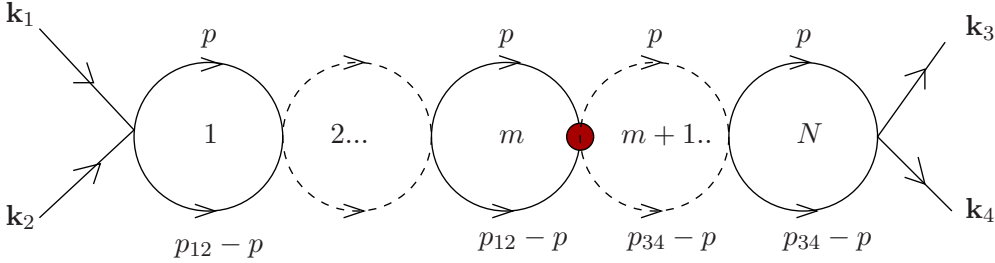} 
\end{center}
\caption{Multi-loop diagrams contributing to the Cooper pair potential $V$.   }  
\vspace{-2mm}
\label{loops} 
\end{figure}

In quantum field theory,   the above matrix element of operators,  in this case  $\CH_{\rm int}$,  is generally  referred to as
a form-factor.   Since there is no integration over time,  this form-factor does not conserve
energy,  i.e. there is no overall $\delta$-function equating $\omega_\kvec $ to
$\omega_{\kvec'}$.     The form-factor can be calculated using Feynman diagrams as follows. 
More generally consider the form-factor  
$\langle \kvec_3  \up,  \kvec_4 \down |   \CH_{\rm int} (\rvec )  |  \kvec_1  \up,   \kvec_2 \down \rangle$.
Represent the interaction vertex with two  incoming arrows for the annihilation operator fields
$\psi_{\up, \down}$ and two outgoing arrows for the creation fields $\psi^\dagger_{\up, \down}$.    
Furthermore,  let such a vertex with a ``node''  $\bullet$ represent the operator $\CH_{\rm int}$.    
Then $V$ is given by the sum over diagrams shown in Figure \ref{loops},  where
$p = (\omega , \kvec )$ represents energy-momentum.

\begin{figure}[htb] 
\begin{center}
\hspace{-15mm} 
\psfrag{0}{$=0$}
\includegraphics[width=4cm]{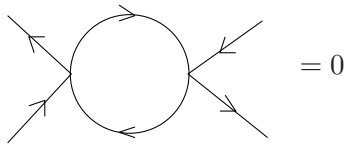} 
\end{center}
\caption{Diagrams with closed loops, i.e. arrows circulating in the 
same direction, vanish.  }  
\vspace{-2mm}
\label{loopzero} 
\end{figure}

  Momentum is conserved at each vertex,  
however energy is not conserved at the vertex with a node.    
There is  actually no fermionic minus sign associated with each loop since the arrows do not form a {\it closed } loop.
Diagrams with a closed loop,  such as in Figure \ref{loopzero},  are zero because the integration 
over energy $\omega$ inside the loop has poles in the integrand that are either both in the upper
or lower half-plane,  so that the contour can be closed at infinity without picking up residues.  In other 
words,  there is no ``crossing-symmetry''  as in relativistic theories.    
(For the contrary,  see the non-zero  loop integral $\CL$ below.)    This fact,  which is unique to non-relativistic theories,  allows us to calculate the kernel $V$ exactly.    At order  $g^{N+1}$,  
specializing to Cooper pairs $\kvec_1 = - \kvec_2 = \kvec$ and 
$\kvec_3 = - \kvec_4 = \kvec'$,    the diagram
in Figure \ref{loops}  factorizes and contributes 
\beq
\label{gN1}
(-i g/2)^{N+1}  \inv{2^N}  \frac{m! (N-m)!}{N!}  \CL(\kvec)^m  \CL (\kvec' )^{N-m}  
\eeq
where $\CL$  is a 1-loop integral
\beq
\label{curlyL}
\CL (\kvec )  =  \int  \frac{ d \omega  d^2 \pvec }{(2\pi)^3}   
\(\frac{i}{ \omega - \omega_\pvec + i \ep } \)
\( \frac{i}{ E_{12} - \omega - \omega_{\kvec_{12} - \pvec} + i \ep}  \)
\eeq
where $E_{12}$ and $\kvec_{12}$ are the total incoming energy and momentum, 
i.e. $E_{12} =  \omega_{\kvec_1 }  + \omega_{\kvec_2 }  = 2 \omega_\kvec$ 
and $\kvec_{12}  = \kvec_1 + \kvec_2 = 0$.   
As usual, $\ep$ is infinitestimally small and positive.   
The extra $1/2^N$ is due to the over-counting by allowing each loop to be $\CL (\kvec)$ 
or $\CL (\kvec' )$.    Finally,  summing over  $m, N$ gives 
\beq
\label{Vfinal} 
V( \kvec , \kvec' ) =   \frac{g/2}{1 + ig(\CL (\kvec) + \CL (\kvec' ))/4}
\eeq
It is important to note that the above $V$ is exact and has a smooth $g\to \infty$ limit, 
i.e. it allows an expansion in the inverse coupling $t/U$, 
so is in a sense non-perturbative.    One may be concerned that we  formally 
summed a geometric series
that potentially does not converge.   In answer to this,   there are certainly regions 
where $\CL$ is small enough that the series converges.   Also,  this  summation is known
to give the correct, exact, S-matrix for non-relativisitic quantum gases,  and this S-matrix
has all of the right properties in the strongly coupled unitary limit\cite{PyeTon2},  
namely,   it gives the correct diverging scattering length at the renormalization group fixed point,
and the bound state.      The only difference here is that the the kinetic energy $\kvec^2/2m$ 
is replaced with $\omega_\kvec$ for the Hubbard model,  which  does not affect 
these arguments.

The  $\omega$-integral can be performed  by deforming the contour to infinity,  giving
$\CL (\kvec )  =  i  \int  d^2 \kvec / 8 \pi^2 (  \omega_{\kvec} - \omega_{\pvec}  + i \ep )$.
Note $\CL$ is imaginary as  $\ep \to 0$,   thus in the formula (\ref{Vfinal}),  
$\CL$ is really the imaginary part of $\CL$ as $\ep \to 0$ such that $V$ is real. 
Then,   integral over $\pvec$ can be performed analytically\cite{SHubbard}:
\beq
\label{CLanalytic}
\CL (\kvec )  = \inv{\pi} 
\(   \frac{  (\omega_\kvec - 4 t' )}{(\omega_\kvec + 4 + 4t')(\omega_\kvec (4 - \omega_\kvec ) + 
16 t' (t'-1)} 
\)^{1/2}
\, 
K\(  \frac{ 16(\omega_\kvec t' -1) }{(\omega_\kvec + 4 t' )^2   -16}  \) 
\eeq
with $\omega_\kvec \to \omega_\kvec + i \ep$,  where $K$ is the complete elliptic integral of
the first kind.    
Note that the momentum dependence of the kernel only enters through the variables 
$\omega_\kvec$,  i.e. $V(\kvec , \kvec' ) = V(\omega_\kvec ,  \omega_{\kvec'})$.

\begin{figure}[htb] 
\begin{center}
\hspace{-15mm} 
\psfrag{x}{$\omega$}
\psfrag{y}{$V$}
\psfrag{glow}{$g=10$}
\psfrag{ghigh}{$g=26$}
\includegraphics[width=10cm]{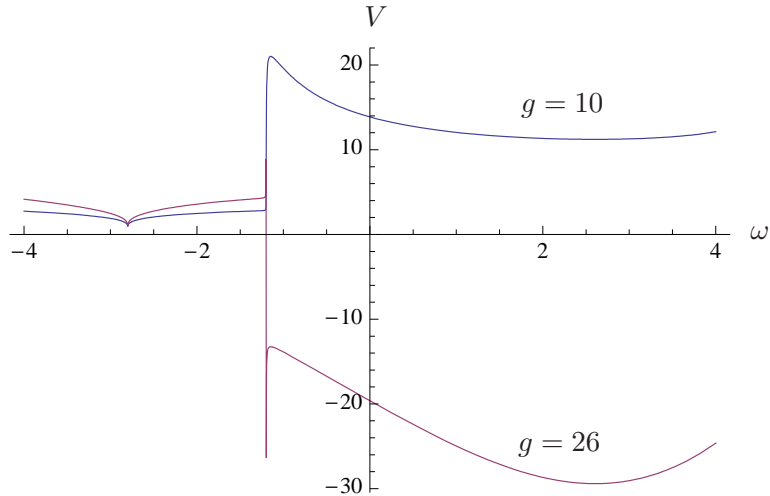} 
\end{center}
\caption{Plot of the  Cooper pair potential $V(\omega, \omega )$ for $g=10, 26$ and  $t' = -0.3$.
(Colored photos on-line).}  
\vspace{-2mm}
\label{V1026} 
\end{figure}

\begin{figure}[htb] 
\begin{center}
\hspace{-15mm} 
\includegraphics[width=7cm]{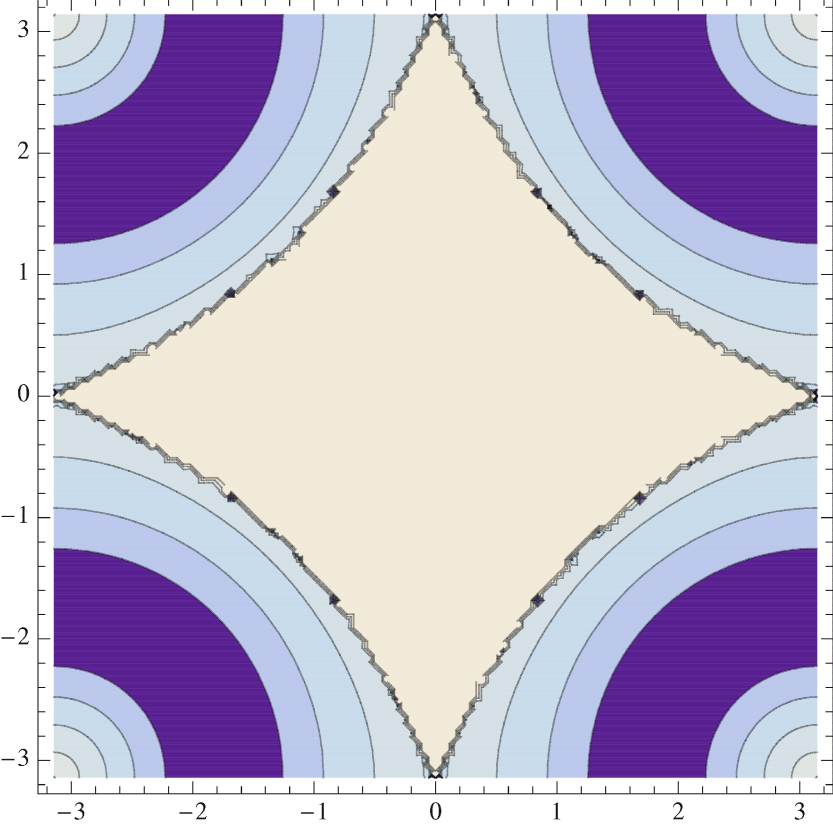} 
\end{center}
\caption{Contour plot of the  Cooper pair potential $V(\omega_\kvec, \omega_\kvec )$ for $g=26$ and  $t' = -0.3$ in the first  Brillouin zone,  with axes $k_x , k_y$.  (All subsequent contour plots in the Brillouin zone follow the same conventions.)      In the central light  region,  the interactions are repulsive, whereas in the 
colored regions attractive.  Values of $V$ can be inferred from Figure \ref{V1026}. 
(Color online).}  
\vspace{-2mm}
\label{Vneg26} 
\end{figure}

Non-zero solutions to the gap equation possibly signifying superconductivity can only arise if the
effective interactions are attractive,  i.e. if the kernel $V$  is negative.    For $g$ small and positive,
the effective coupling $V$  remains repulsive.    However,   as pointed out in \cite{SHubbard}, 
for $g$ large enough,  $V$ can become negative in certain regions of the Brillouin zone.    
Since we are interested in a small band of energies near the Fermi surface,  $V(\omega, \omega') $
with $\omega' = \omega$ is a suitable  probe of these attractive regions.    In Figure \ref{V1026} 
we plot this $V$  for $g=10$ and $26$ at fixed $t'=-0.3$.     One sees that for the smaller $g$,  
$V$ is everywhere positive,  however for larger $g$ it  flips sign.   As explained in more detail in
\cite{SHubbard},   this feature is reminiscent of  what occurs near the fixed point of 
quantum gases in the unitary limit,  where for the same analytic reasons,  the effective interactions
can be either attractive or repulsive depending on which side of the fixed point 
of  the BEC/BCS crossover\cite{PyeTon2}.  
     Using the formula
(\ref{CLanalytic}),   one can show that there is a region of negative $V$ for $g> 13.2$.   
This minimal value of $g$ depends on $t'$ and  this dependence was studied in \cite{SHubbard}
based on the formula (\ref{CLanalytic}).    
Around $g=13-15$,  the attractive region is a narrow band\cite{SHubbard}.    
It will  also be instructive to view a contour plot of $V$ in the first Brillouin zone,  see Figure
\ref{Vneg26}.

The main effect of a non-zero $t'$ is the following.   For $g$ large enough,  $V(\omega, \omega)$ 
becomes negative for $\omega > 4 t'$.   Thus when $t'$ is negative and $|t'|$ large,   
attractive interactions exist deeper inside the half-filled Fermi surface.  If superconductivity 
indeed arises from these attractive interactions,  then a non-zero $t'$ can play a significant role, 
otherwise the attractive interactions only exist too close to the vicinity of the half-filled   Fermi surface 
where it has to compete with the known Mott-insulator phase.    There is actually some evidence
that superconductivity does not exist for $t'=0$\cite{tprimedata}.


\section{The Fermi surfaces as a function of doping}

We will utilize the approach to the thermodynamics of particles developed in 
\cite{PyeTon,SHubbard},  which is based on a self-consistent re-summation of
the exact 2-body scattering.     The occupation numbers are parameterized by two 
pseudo-energies  $\vep_{\up, \down}  (\kvec )$, which satisfy 2 coupled integral equations
with a kernel related to the scattering of spin up with spin down.   For equal chemical potentials, 
due to the SU(2) symmetry, 
$\vep_\up = \vep_\down$,  and both occupation numbers are given  by 
\beq
\label{fvep}
f(\kvec) =  \inv{ \rme^{\vep(\kvec)/T} +1 } 
\eeq
where $\vep$ satisfies the  single  integral equation:
\beq
\label{inteq}
\vep (\kvec )  = \omega_\kvec - \mu  -  \int  \frac{d^2 \kvec' }{\2pi2} \, G(\kvec, \kvec' )  
\inv{  \rme^{\vep (\kvec')/T}  +1 }  
\eeq

The kernel $G$ is related to the logarithm of the 2-body S-matrix,  and is built from the same
ingredients as the kernel $V$ in the gap equation,  since it  also involves a sum of Feynman diagrams
of the kind shown in Figure \ref{loops}.   It is somewhat more complicated than the pair potential $V$
since the total incoming momentum is not zero.  
  Namely,  consider the same loop integral as in the previous section
but with $\kvec_1 + \kvec_2 \neq 0$:
\beq
\label{Ldef}  
L (\kvec_1 , \kvec_2 )  =  i   \int  \frac{d^2 \pvec} {\2pi2} ~
\inv{ \omega_{\kvec_1}  + \omega_{\kvec_2}  - \omega_\pvec  -  \omega_{\kvec_1 + \kvec_2 - \pvec} 
+ 2 i \ep}
 ~~~\equiv ~~  \CI  + i \gamma 
\eeq
where $\CI$ and $\gamma$ are defined to be real.   
Then the kernel takes the following form 
\beq
\label{Gdef}
G = -\frac{i}{2\CI} \log \(  \frac{1/g_R  - i \CI/2}{1/g_R + i \CI/2}  \)
\eeq
where the renormalized coupling is 
$g_R  = g/(1 - g \gamma/2)$.
 (We are not displaying the momentum dependence;
it is implicit that $G= G(\kvec=\kvec_1 , \kvec' = \kvec_2 )$.)   
The renormalized coupling is related to the gap equation kernel of the last section 
as follows:  $g_R (\kvec, -\kvec )  =   2V(\kvec, \kvec )$.   
The quantity $\CI$ represents the phase space available for 2-body scattering.
The argument of the $\log$ is the exact 2-body S-matrix.   

We define the hole doping $h$ as the number of holes per plaquette,  which is related to the
density  $n$ as follows:
\beq
\label{hdef}
n  =  2  \int  \frac{d^2 \kvec }{\2pi2} ~ \inv{ \rme^{\vep (\kvec ) /T} +1 }   =  \frac{1-h}{a^2}
\eeq
where $a$ is the lattice spacing.      

\begin{figure}[htb] 
\begin{center}
\hspace{-15mm} 
\psfrag{x}{$\mu$}
\psfrag{y}{$h$}
\includegraphics[width=10cm]{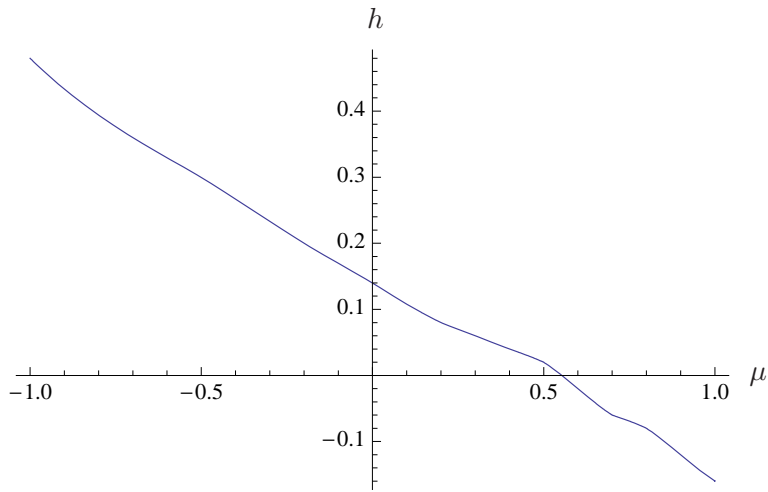} 
\end{center}
\caption{Hole doping $h$ as a function of chemical potential $\mu$ at
the reference temperature $T_0 = 0.1$.  ($g=26, t'=-0.3$.)}  
\vspace{-2mm}
\label{muofh} 
\end{figure}

The integral equation (\ref{inteq}) was solved numerically using an iterative procedure,
as explained in \cite{SHubbard}.   
The  solution for the pseudo-energy yields 
the relation between the chemical potential $\mu$ and the hole doping $h$.    For a given $h$,
$\mu (h)$ of course depends on temperature,  but only weakly\cite{SHubbard}.   
For the subsequent analysis we determine $\mu (h)$ at the low reference temperature 
$T_0 = 0.1$.    The result is shown 
in Figure \ref{muofh}.

As described in \cite{SHubbard},  for low enough $T$,    there are regions of
$\mu, T$ where there are no solutions to the integral equation at low enough $T$.   
Regions of the non-existence of solutions  are very similar to that shown in Figure 8  in\cite{SHubbard}, 
For hole dopings $0<h<0.25$,  there are no solutions for temperatures  $T_c$ 
in the range $0.02< T_c < 0.07$.   
(This is why we chose the reference temperature $T_0=0.1$ to be above these  potential 
transition temperatures.)  
It was suggested in \cite{SHubbard}   that  the non-existence of solutions 
 could represent  an instability toward the formation of a new phase,  however the
nature of these new phases cannot be determined based on what we have done so far;  
one needs a complementary bottom up approach that is based on the zero temperature
ground state.   This is the subject of the next section.   
As we will see,   the critical temperatures inferred from  this thermodynamic analysis 
are consistent with the critical tempertures computed from the gap equation
in the next section.

\begin{figure}[htb] 
\begin{center}
\hspace{-15mm} 
\includegraphics[width=10cm]{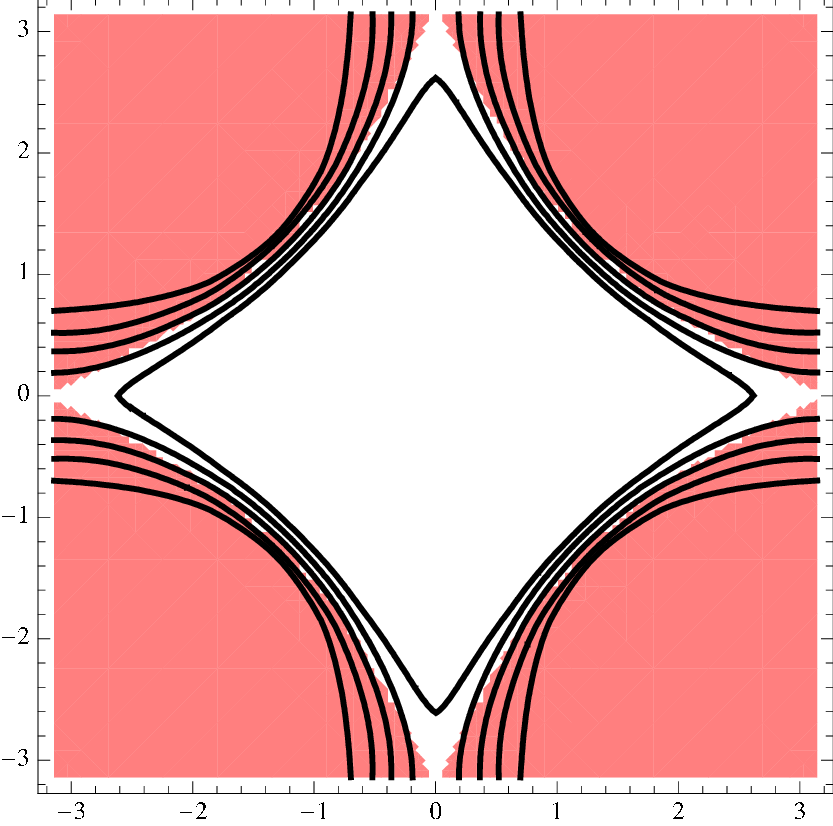}
\end{center}
\caption{Calculated Fermi surfaces for hole doping 
 $0  <h < 0.4$ in steps of $0.1$,  with $h=0.4$ the innermost curve.  
The grey (pink on-line)   region corresponds to the attractive region for $g=26, t'=-0.3$ based on the 
Cooper pair potential $V$.}
\vspace{-2mm}
\label{Atband} 
\end{figure}

The Fermi surfaces are the contours $\CS_F (\mu )$ that are solutions to 
$\vep (\kvec, \mu) =0$,  where  $\mu$ depends on hole doping as in Figure \ref{muofh}.  
The pseudo-energy $\vep$ thus represents the quasi-particle dispersion relation 
in the normal state.   
These Fermi surfaces are shown in Figure \ref{Atband}  for  $0<h < 0.4$.   
These calculated Fermi surfaces are in reasonably good agreement with 
experiments.  Comparison with Figure \ref{Atbandxifit}, which is based on
a phenomenological fit to the data,  suggests that one needs to include additional hopping
terms in the bare hamiltonian,  such as next-to-next neighbor.  
We will refer to wave-vectors $\kvec$ that point to $(\pi, \pi)$ (and $90^\circ$ rotations thereof) 
as being in the nodal direction, 
whereas those pointing to  $(0, \pi)$ as in the anti-nodal direction.   

In the same Figure \ref{Atband}  we also display the region of attractive interactions based on
the Cooper pair potential $V$ calculated in the last section.  
  Before even solving the gap equation,  one can make
some predictions concerning the existence of superconductivity based on the attractive
region of $V$.   Namely,  for high enough hole doping,  about  $h>0.3$,   there is no attractive region
near the Fermi surface,  and thus no superconductivity.      It is also clear from Figure \ref{Atband} 
that the regions of the Fermi surface in the anti-nodal direction are the most important since 
this is the direction with  the greatest overlap with the attractive region.   As we will show in the next section,
this feature is primarily responsible for the anisotropy of the gap $\Delta (\kvec )$,  and explains why
the gap is zero in the nodal direction,  at least for moderately high doping $h>0.1$.

\section{Solutions to the gap equation}

The gap $\Delta (\kvec)$ is  defined and measured for $\kvec$ along the Fermi surface.    
   However near half-filling,  there is no $\kvec$ from
the origin that intersects the Fermi surface in the anti-nodal direction;  see  Figure \ref{Atband}.    
Consequently,   polar plots of $\Delta (\kvec )$ for $\kvec$ originating from  the center of the 
Brillouin zone,  and covering all of it,  are potentially misleading,   since $\Delta (\kvec)$ is not
well-defined in the anti-nodal directions.    In fact,  this also implies that $\Delta$ cannot be
strictly d-wave as defined from the origin, 
  for example,  it cannot be of the simple form   $\Delta \propto | \cos k_x  - \cos k_y|$,  where
  $\kvec$ is measured from the origin.       
Thus,  in the first quadrant of the Brillouin zone,  
 it is more convenient to work with the vector $\kappavec$  originating from the node 
$(\pi, \pi)$,  i.e.  $\kvec =  \kvec_{\pi, \pi}  + \kappavec$,  where
$\kvec_{\pi , \pi}  = (\pi , \pi)$.

In the integral gap equation (\ref{gapeqn}),  one must integrate over a narrow band around the 
Fermi surface $\CS_F (\mu )$.    
Let $\theta$ be the angle of $\kappavec$ relative to the horizontal line through the $(\pi, \pi )$ node:
$\kappavec =  - \kappa ( \cos (\theta) \,    \ihat  +  \sin (\theta)  \,  \jhat )$.   If  the integration is over
a narrow band  $\delta \CS_F$ of width $2 \, \cutoff$ around the Fermi surface,  then the 
gap equation  
takes the following form in the first quadrant:
\beq
\label{gapeqnnodal}
\Delta (\kappa, \theta )  = 
 -\inv{8 \pi^2}  \int_0^{\pi/2}   d\theta'  \int_{\kappa_F (\theta') - \cutoff}^{\kappa_F (\theta') 
+ \cutoff}  d \kappa'   \kappa'  ~   V_{-} ( \kappa, \theta; \kappa', \theta' )   
\frac{\Delta (\kappa', \theta' ) }{E(\kappa', \theta' )}   \tanh \(  \frac{E(\kappa', \theta' )}{2T} \) 
\eeq
where 
$E =   + \sqrt{ \xi^2 + \Delta^2 }$, 
and  $\kappa_F (\theta)$ is the length of $\kappavec$ along the Fermi surface.   
Here  $V_-$ is the negative (attractive) part of $V$ since only when $V$ is negative are there
solutions;   the repulsive parts of $V$ are already incorporated in determining the 
Fermi surfaces.        The doping $h$ dependence of the 
above equation is implicit in  $\kappa_F (\theta)$.

It remains to determine the cut-off $\cutoff$. 
There is some arbitariness in the choice of $\cutoff$,  and this is the weakest aspect of our 
calculation since $T_c$  certainly depends on it,  just as $T_c$ depends on the Debye frequency 
in ordinary superconductors.      It could
be viewed as a free parameter that needs to be fit to the data. 
Or,  one could carry out a sophisticated renormalization group analysis requiring $\Delta$ to
be independent of $\cutoff$,  but this is at the expense of introducing an arbitrary scale that
has to be fit to experiments.     Instead,  
   we found  the following
choice  to be physically meaningful and well-motivated.   
As in the BCS theory,  $\cutoff$ should be related to 
the properties of the potential $V$ itself,  namely the region in which it is attractive relative to 
the Fermi surface\footnote{In the BCS theory,  the cut-off is the Debye frequency $\omega_c$,
and is a measure of the width of the attractive region.}.  
    Define  $\kappa_V$ such that $V$ is attractive for $\kappa < \kappa_V$
in the nodal direction.   For $g=26,  t'=-0.3$,  $\kappa_V \approx 2.7$.     We then take 
$\cutoff$  as a measure of the distance of the Fermi surface  to the edge of the  region of attractive interactions in the nodal direction:   
$\cutoff = |\kappa_V - \kappa_F (\pi/4)|$.    From the Figure \ref{Atband},  one 
sees that $\cutoff$ is quite small, so that our choice does correspond to a narrow band. 
  For the Fermi surfaces computed in the last section,
$\cutoff \approx 0.04$ for doping $h=0.15$ and this  value will be  used in the subsequent analysis.

The function $\xi (\kvec )$ in the gap equation represents the quasi-particle dispersion relation 
of the normal state,  and is zero along the Fermi surface.   We wish to carry out a self-contained
calculation,  thus we equate $\xi$ with the pseudo-energy $\vep$ of the last section.   
On the other hand,  one can infer $\xi$ from experiments,  and we will repeat the analysis 
with such a $\xi$ in the next section.

A measure of the validity of the BCS approximation is the combination of the effective coupling $V$
and the phase space integration in the gap equation,  i.e. the parameter 
 $ v = -\kappa_F \cutoff V/8\pi$.    From Figure 
\ref{V1026},   $V\approx -20$,  and also $\kappa_F \approx 2.7,   \cutoff\approx 0.04$.  
This gives $v\approx 0.05$,  which appears to be sufficiently small for the BCS approximation to
be valid.

\begin{figure}[htb] 
\begin{center}
\hspace{-15mm} 
\psfrag{x}{$\theta ({\rm degrees})$}
\psfrag{y}{$\Delta$}
\psfrag{hlow}{$h=0.10$}
\psfrag{hhigh}{$h=0.20$}
\includegraphics[width=10cm]{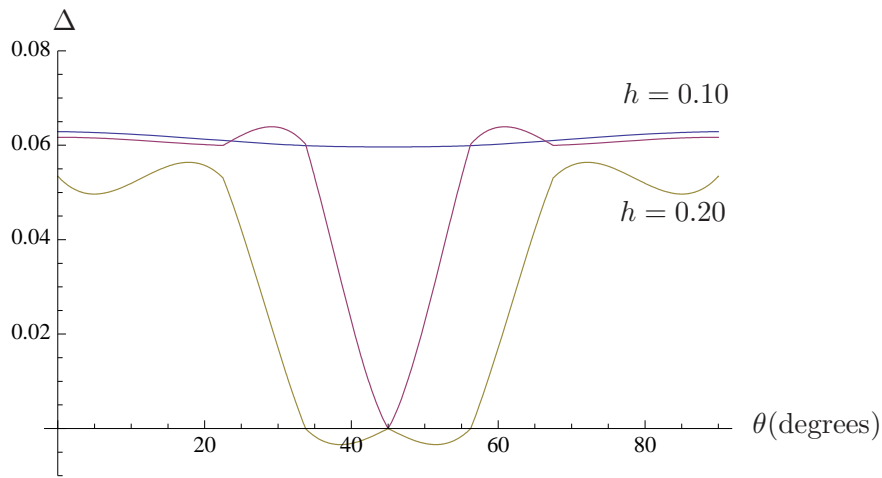} 
\end{center}
\caption{The zero temperature gap $\Delta$ as a function of  the Fermi surface angle $\theta$  
for hole doping  $h = 0.10, 0.15, 0.20$.}  
\vspace{-2mm}
\label{Delta3h} 
\end{figure}

We solved the gap equation numerically by discretizing the integrals and solving the 
resulting set of coupled non-linear equations with Mathematica.        In Figure \ref{Delta3h} 
we plot the zero temperature  gap as a function of  the Fermi surface angle $\theta$ for 3 different hole dopings.   
One sees that $\Delta (\kvec )$ is highly anisotropic in the Brillouin zone:   it is largest in
the anti-nodal directions,  and zero in the nodal direction for high enough doping.    At low doping,  the anisotropy
is less pronounced.      This property can be traced to
the detailed shape of the Fermi surface in comparison to the region of attractive interactions 
displayed in Figure \ref{Atband}.    More specifically,  for high enough doping,   the Fermi surface
does not intersect the region of negative $V$ in the nodal direction,   however it always does in
the anti-nodal direction.      This effect is even more pronounced for the experimentally determined
Fermi surfaces,  as will be described in the next section.        

Our solutions to the gap equation are not exactly d-wave,  more specifically,  are not proportional 
to $|\cos \kappa_x  -   \cos \kappa_y|$,  which is approximately   $\kappa_F |\cos 2\theta| /2$,
for the Fermi surface  approximated as a  circle of radius $\kappa_F$.     Experimental data 
indicates a gap closer to the d-wave form,  however some data  does show a tendency for it to flatten 
out in the nodal direction,  as in our solutions.  
However the precise shape of the gap as a function of $\theta$ will change if the cut-off
$\cutoff$  is made to depend on $\theta$,   instead of a simple constant as we have done here.   
Some features in Figure \ref{Delta3h}  are reflected in the data\cite{Seamus},  in particular,
the central region around $\theta = 45^\circ$ where the gap is smallest widens with increasing
doping.   We also wish to point out that our Figure \ref{Atband}
is  suggestive of an observation made in \cite{Seamus}:   the  Bogoliubov quasiparticle
interference,  indicative of the existence of Cooper pairs,  disappears along the diagonal line
connecting the two anti-nodes $(0, \pi)$ and $(\pi, 0)$. (See Figure 3 in \cite{Seamus}.)
  Interestingly,   this diagonal line
is very close to the contour that separates attractive from repulsive regions of the pair potential,  however
we are unable to make a direct connection at present.   These observations are 
more pronounced in Figures  \ref{Gapsxi} and  \ref{Atbandxifit}.

\begin{figure}[htb] 
\begin{center}
\hspace{-15mm} 
\psfrag{x}{$T$}
\psfrag{y}{$\Delta$}
\includegraphics[width=10cm]{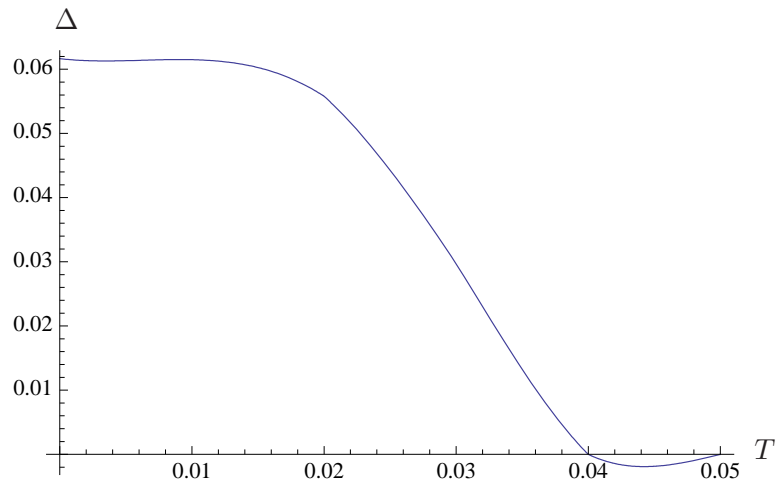} 
\end{center}
\caption{The gap $\Delta$  in the anti-nodal direction as a function of $T$ for hole doping $h=0.15$.}  
\vspace{-2mm}
\label{DeltaofTh15} 
\end{figure}

\begin{figure}[htb] 
\begin{center}
\hspace{-15mm} 
\psfrag{x}{$h$}
\psfrag{y}{$\Delta, T_c$}
\psfrag{delta}{$\Delta$}
\psfrag{Tc}{$T_c$}
\includegraphics[width=10cm]{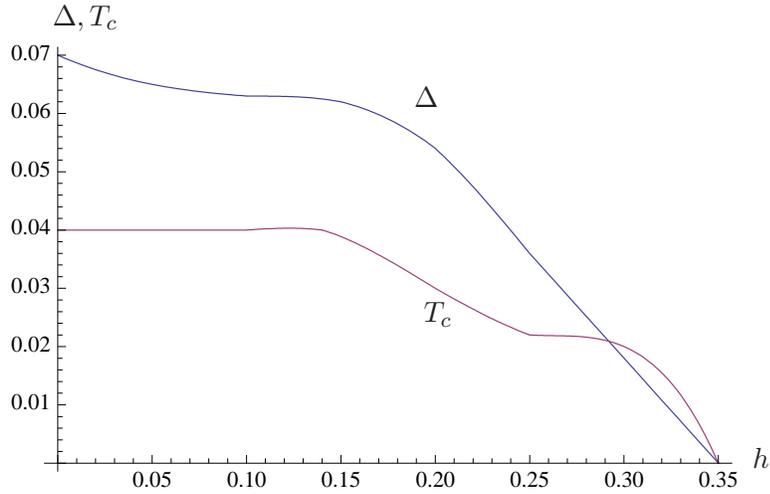} 
\end{center}
\caption{The zero temperature gap $\Delta$ in the anti-nodal direction and $T_c$, in units of $t$,  as a
function of doping $h$.}  
\vspace{-2mm}
\label{DeltaAndTc} 
\end{figure}

The gap in the anti-nodal direction  is plotted as a function of temperature for $h=0.15$ in Figure \ref{DeltaofTh15}.    
Where it goes to zero defines $T_c$,  in this case $T_c \approx 0.04$.     Our results for the
zero temperature gap in the anti-nodal direction and $T_c$  for various doping are summarized in 
the table below.  
Figure  \ref{DeltaAndTc}  plots both  the zero temperature gap and $T_c$ as a function of doping.   
Moving down from the overdoped side,  the maximum $T_c$ occurs first at $h=0.15$,  in 
good agreement with experiments.   
 For $t=3000K$,  one obtains $\Delta = 19 meV$ and $T_c = 120K$ 
at $h=0.15$,  compared with the experimental values $\Delta = 30 meV$ and 
  $T_c = 90K$ for BSCO.  

\begin{center}
\begin{tabular}{|c|c|c|}
\hline\hline
$~{\rm hole-doping } ~h~$     &   $~\Delta_{\rm anti-nodal}~$   &  $~~~~T_c~~~~$    \\
\hline\hline 
$0.0$  & $ 0.070   $ &  $0.04$           \\
$0.05$  & $ 0.065  $ &  $0.04$           \\
$0.10$  & $ 0.063  $ &  $0.04$           \\
$0.15$  & $ 0.062   $ &  $0.04$           \\
$0.20$  & $ 0.054   $ &  $0.03$           \\
$0.25$  & $ 0.036   $ &  $0.022$           \\
$0.30$  & $ 0.018  $ &  $0.02$           \\
$0.35$  & $ 0.0   $ &$0.0$           \\
\hline\hline 
\end{tabular}
\label{table1}
\end{center}

As explained in the last section,   there is a succinct reason 
for why there is no superconductivity at high enough doping,  roughly $h>0.35$, 
since beyond this,   no part of the Fermi surface overlaps with  $V_-$.    
See Figure \ref{Atband}.   A slightly lower value of $h>0.3$ is more typical in experiments.  
However there is no mechanism in the gap equation to turn off the gap at low enough doping, 
and it continues to increase all the way down to zero doping.  
Superconductivity would turn off if $g$ were smaller,  namely around $g=15$,  since
in this case the region of attractive interactions is a narrow band,  and at low enough doping
the Fermi surface does not overlap with it\cite{SHubbard}.     However if eq. (\ref{gdef}) is 
accurate,   $g$ is roughly twice as large,   so this does not account for the disappearance
of superconductivity at low doping.     Although this may appear problematic when one compares
with the usual phase diagram of the cuprates,   there is growing experimental evidence that
this is actually  the correct behavior\cite{Seamus,Chatterjee}.   Namely,  it has recently been 
found that the d-wave superconductivity gap evolves smoothly into a d-wave pseudogap 
whose magnitude continues to increase to arbitrarily low doping,  where superconductivity 
is not present.    
In other words,  
the so-called pseudo-gap energy scale $T^*$  may be  the continuation of the superconducting gap,
{\it had there been no other other mechanisms to destroy it}.    As stated explicitly in 
\cite{Chatterjee},  these results are inconsistent with 2-gap scenarios.    One check of this is that if
one identifies $T^*$ with $\Delta$,   then $T^*= 195K$ for $t=3000K$ at  doping $h=0.05$,  again
in reasonable agreement with experiments.  
This suggests that  on the underdoped side,  
superconductivity is  perhaps  destroyed by competition with other orders,  
presumably anti-ferromagnetic,  bringing $T_c$ to zero,  even though the gap $\Delta$ is
still physically present.     
It could also be destroyed by phase decoherence,  as suggested in\cite{Tesanovic,Nayak}.   
These effects are of course not implemented in our gap equation,  and 
it is beyond the scope of this paper to address this,  for example by comparing the free energies for
the competing orders or to study to phase fluctuations.


\section{Solutions to the gap equation for phenomenologically determined Fermi surfaces}

In this section we repeat the analysis of solutions of the gap equation with the normal state quasi-particle 
dispersion relation $\xi (\kvec )$  determined from experimental data,   but with the same 
pair potential $V$ computed in section III.     
There is extensive data on the Fermi surfaces for the compound 
${\rm Bi_2 Sr_2 Ca Cu_2 0_{8+\delta}}$.    A tight-binding fit to the data was performed in
\cite{Norman} based on the data in \cite{Ding}.  
   The result is that the Fermi surfaces are the contours $\xi (\kvec ) =0$  for the
following function:
\barray
\nonumber 
\xi (\kvec )   &=&   -\hat{\mu}  - 2 \( \cos k_x + \cos k_y \) +  0.6513 \cos k_x  \cos  k_y  
 - 0.4455 \( \cos 2k_x  + \cos 2 k_y  \)  
 \\  
 \label{xifit}
  &~&  ~~~
    - 0.1716  \( \cos 2 k_x  \cos  k_y  + \cos k_x  \cos 2 k_y \) 
  + 
 0.6357 \cos 2 k_x  \cos  2 k_y  
 \earray
 We have rescaled  the result for $\xi$  in \cite{Norman}  so that $t=1$.      The parameter $\hat{\mu}$ serves as a 
 renormalized chemical
 potential.
For hole dopings $h=0.07, 0.08$, $ 0.14, 0.17, 0.19$,   $\hat{\mu}  =  -0.0500, -0.178, 
-0.503, -0.688, -0.809$  respectively.  

\begin{figure}[htb] 
\begin{center}
\hspace{-15mm} 
\includegraphics[width=10cm]{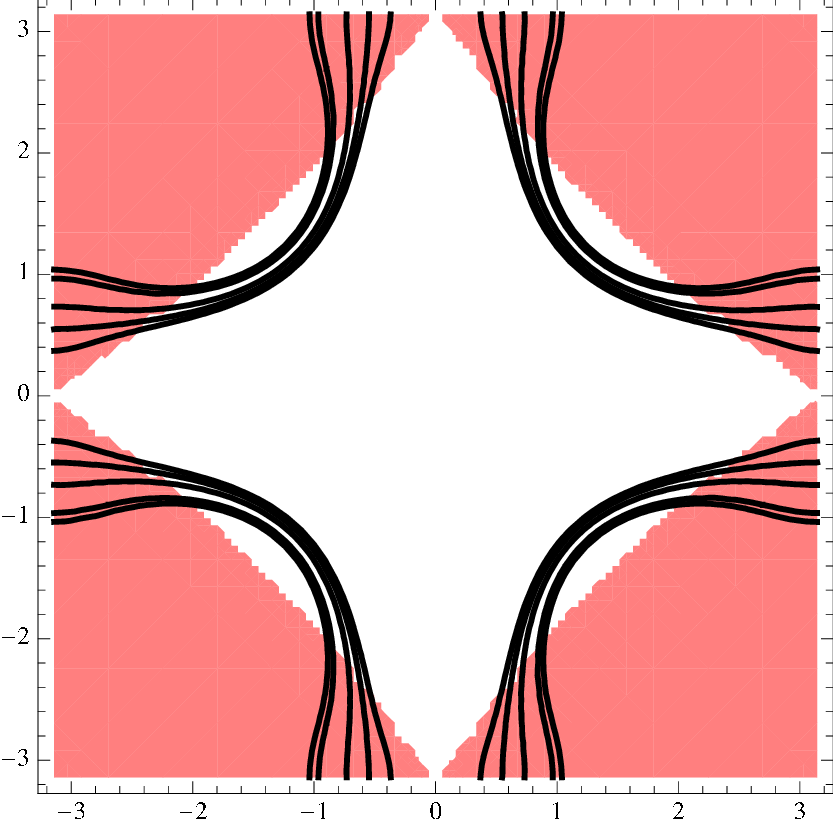}
\end{center}
\caption{Tight-binding fit to the Fermi surfaces for ${\rm Bi_2 Sr_2 Ca Cu_2 0_{8+\delta}}$ 
at dopings  $0.07, 0.08, 0.14, 0.17, 0.19$,   eq. (\ref{xifit}).   
The grey (pink online)  region corresponds to the attractive region for $g=26, t'=-0.163$ based on the 
Cooper pair potential $V$.}
\vspace{-2mm}
\label{Atbandxifit} 
\end{figure}

Let us assume that the underlying Hubbard hamiltonian has the same $U/t =13$, i.e.  $g=26$,  and 
 still only has nearest and next-nearest neighbor
hopping parameters $t$ and $t'$,  where from eq. (\ref{xifit}) one reads off  $t'= -0.163$.     
The other terms in eq. (\ref{xifit}) should be  viewed as generated by the interactions,    for example 
by equations such as in section IV for the pseudo-energy $\vep$.   
In Figure \ref{Atbandxifit}  we  display the resulting Fermi surfaces against the 
region of attractive interactions $V_-$ as computed in section III,  but  with $t'=-0.163$.  
In comparison with Figure \ref{Atband},  one sees that the effect that leads to the anisotropy of
the gap is more pronounced:   the Fermi surfaces are pulled away from the attractive region in the
nodal direction in a stronger  manner,  which implies that the gap will continue to
 be zero in the
nodal direction for lower values of $h$ in comparison to the last section.

\begin{figure}[htb] 
\begin{center}
\hspace{-15mm} 
\psfrag{x}{$\theta ({\rm degrees})$}
\psfrag{y}{$\Delta$}
\psfrag{hlow}{$h=0.07$}
\psfrag{hhigh}{$h=0.14$}
\includegraphics[width=10cm]{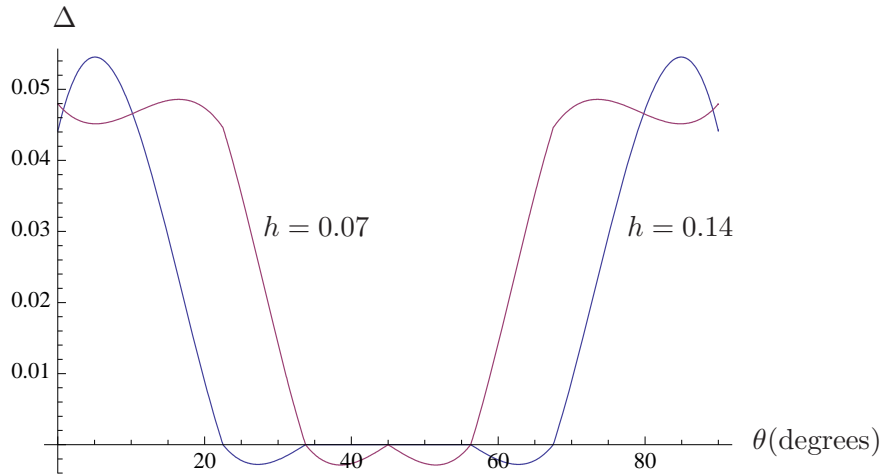} 
\end{center}
\caption{The zero temperature gap $\Delta$ as a function of $\theta$ for hole doping  $h = 0.07, 0.14$ 
based on the phenomenological $\xi$ eq. (\ref{xifit}).}  
\vspace{-2mm}
\label{Gapsxi} 
\end{figure}

In Figure \ref{Gapsxi}  we plot the solution to the gap equation for $h=0.07,  0.14$.    
The same cut-off $\cutoff=0.04$ as in the last section was used.   
For the reasons stated above,   the gap is zero for a wider region centered at $\theta = 45^\circ$.   
Plots of the gap as a function of temperature are very similar to those of the last section,
and lead to a slightly lower $T_c$,  i.e. $T_c \approx 0.03$ at optimal doping $h=0.14$.

\section{Conclusions}

The central proposal of this work is that quantum loop corrections to the Cooper pair potential,  
as computed here  in the two-dimensional Hubbard model,   could be responsible for the effectively attractive interactions near the Fermi surface that lead to 
the phenomenon of high $T_c$ superconductivity.     The validity of this idea
is easily explored,  since the pair potential can be calculated exactly, 
and the results favor our proposal.   
We showed that the resulting analysis of the solutions to the superconducting gap equation 
leads  to definite predictions  for the anisotropy of the gap,  its magnitude,  and $T_c$,  
all in reasonably good agreement with experiments.   We explained 
in a clear manner why superconductivity disappears at high enough doping;     in our calculation 
$h>0.35$.     

We found that the non-zero solutions to the gap equation continue undiminished on the underdoped side,  all the way down to zero doping,  and this appears to be consistent with recent experimental 
results\cite{Seamus,Chatterjee},  with the interpretation that the d-wave superconductivity gap
evolves smoothly to the d-wave pseudogap.      If the attractive interactions considered in this paper are
indeed responsible for superconductivity,  then this feature suggests two items on the more
speculative side:   

$\bullet$   On the underdoped side,  superconductivity is  perhaps destroyed by competition with the 
anti-ferromagnetic phase which is known to exist at very low doping.    Here one must bear in mind
that we kept only the attractive part of the pair potential for the superconducting gap equation,  
but inside the Fermi surface the interactions are still largely repulsive.  
Another possibility is that it is destroyed by phase decoherence of the gap\cite{Tesanovic,Nayak}.     

$\bullet$   The pseudo-gap energy scale $T^*$  may thus  represent the hypothetical continuation of the
superconducting gap had it not been destroyed by the mechanisms suggested above.    
If one identifies $T^*$ with $\Delta$,   then $T^*= 210K$ for $t=3000K$ at  zero doping,  again
in reasonable agreement with experiments.

If these ideas are correct,   then the emphasis should shift from trying to understand 
 ``doping the Mott insulator'' to its opposite,  that is to say,   understanding how populating the superconducting state can destroy it due to the competing anti-ferromagnetic  order,  
 phase decoherence,  or perhaps something else. 
 This  issue  has been studied experimentally in significant detail\cite{Seamus}.   
The latter approach  may be more tractable if based on the  concrete description of
high $T_c$  superconductivity presented in this paper.

\section{Acknowledgments}

   I would like to thank  Jacob Alldredge,   Seamus Davis,   Eliot Kapit,   Kyle Shen,  and   Henry Tye for discussions.  
   I also wish to thank members of the Centro Brasileiro de Pesquisas F\'isicas in Rio de Janeiro,
   especially Itzhak Roditi,  for their kind hospitality during the completion of this work. 
    This work is supported by the National Science Foundation
under grant number  NSF-PHY-0757868.

\end{document}